# Chemical Evolution of Dwarf Irregular Galaxies – chemodynamical models and the effect of gas infall

Gerhard Hensler[1], Andreas Rieschick[1], Joachim Köppen[1,2,3]

[1] *Institut für Theoretische Physik und Astrophysik, Universität Kiel,
D–24098 Kiel, Germany; email: hensler@astrophysik.uni-kiel.de*
[2] *UMR 7550, Observatoire Astronomique de Strasbourg,
11 rue de l'Universit, F–67000 Strasbourg, France*
[3] *International Space University, Parc d'Innovation,
Blvd. Gonthier d'Andernach, F–67400 Illkirch, France*

**Abstract.**   Because of their low gravitational energies dwarf galaxies are
greatly exposed to energetical influences by the interstellar medium, like
e.g. stellar radiation, winds or explosions, or by their environment. While
the metallicity depletion in dwarf galaxies can be explained in general by
supernova-driven galactic winds, the reason for their low N/O ratios at
low O abundance is not yet completely understood. Stellar yields enrich
the different gas phases with elements that are characteristic for their
stellar progenitors. Gas-phase transitions are necessary for a mixing of
elements, but depend sensitively on the thermal and dynamical state of
the interstellar medium. Models of chemical evolution start usually with
a high N/O ratio at low O abundance according to a metal enrichment by
ancient stellar populations with traditional yields, but can only reproduce
the N/O-O peculiarity by the stepwise element release and mostly by the
application of multiple starbursts in order to account also for a selective
element reduction by galactic winds.

Chemodynamical models of dwarf galaxies, however, demonstrate
that strong evaporation of clouds by the hot supernova gas leads to an
almost perfect mixing of the interstellar gas and a large-scale homoge-
nization of abundance ratios. In addition, even with star-formation self-
regulation these models can successfully account for the observed N/O-O
values in a self-consistent way. Although new stellar yields have been
taken into account which provide mainly secondary N production from
massive stars, a significant discrepancy remains between the chemody-
namical dwarf galaxy models and closed-box models that take the chemo-
dynamical treatment into account. In this paper we therefore discuss to
what amount gas infall is responsible to affect the N/O ratio.

## 1.   Introduction

### 1.1.   Morphologies and Star-formation Epochs of Dwarf Galaxies

Dwarf galaxies (DGs) present a wide variety of morphological types.  Their
structural and chemical properties differ from those of giant galaxies.  In addi-tion, low-mass galaxies seem to form at all cosmological epochs and by different processes and from different sources. Dwarf elliptical galaxies (dEs) are an extremely common and astrophysically interesting class of galaxy. Most known dEs are found in regions with high galaxy densities, and they are the most numerous of all galaxy types in the cores of nearby galaxy clusters (see e.g., Wirth & Gallagher 1984, Binggeli 1994). Since the bulk of their star formation (SF) occurred in the past (see review by Ferguson & Binggeli 1994), dEs thus are frequently considered to be "stellar fossil" systems. The compact "blue dropouts" (Madau et al. 1996) visible in the Hubble Deep Field (HDF; Williams et al. 1996) seem to reflect the formation epoch of dEs by part. Their number excess compared to the present dE density requires, however, the disappearance of some fraction, caused by three different processes: firstly, disruption by means of vehement evolutionary effects like starbursts (SBs) and subsequent intense type II supernova explosions (SNII) or, secondly, by means of tidal forces and last but not least due to their accretion by massive galaxies.

In galaxies where gas is consumed by astration stellar abundances are predicted to be close to the solar value. On the other hand, the moderate-to-low stellar metallicities in dE and the related dwarf spheroidals (dSphs) which represent the low-mass end of DGs (about 0.1 of solar or less) suggest that extensive gas loss occurred during their evolution by means of SNII driven galactic winds (Larson 1974, Dekel & Silk 1986). Yet many dSph galaxies show not only a significant intermediate-age stellar population (Hodge 1989, Grebel 1997), but also more recent SF events (Smecker et al. 1994, Han et al. 1997) with increasing metallicity, indicating that gas was partly kept in the system. On the other hand, gas infall might also cause the excitation of a new SF episode like in NGC 205 (Welch et al. 1998).

At medium redshifts faint blue galaxies appear (Colless et al. 1990, Lilly et al. 1991) which resemble spectroscopically local compact HII regions with respect to their strong but narrow emission lines, their colors and the M/L ratio (Koo et al. 1995, Guzman et al. 1998, Guzman, this conference). Those extremely high SF rates exist at present also in a special DG type, which are characterized by small sizes, blue centrally bright colours, HII-type emission lines, high gas content and low metallicity: blue compact DGs or also called HII galaxies. Early survey studies e.g. by Haro, Zwicky, and Markarian have drawn attention to this class of objects. In general, gas-rich dwarf irregular galaxies (dIrrs) show a large variety of SF rates from low to extraordinarily high values as in SBs but consist additionally of at least one intermediate-age to old underlying stellar population like in NGC 1569 (Heckman et al. 1995) or NGC 1705 (Meurer et al. 1992). The newly formed stellar associations are often very massive and compact and, therefore, called super star clusters (SCC) which are assumed to evolve to Globular Clusters by aging (Ho & Fillipenko 1996).

The addressed questions why and by what physical process such enormous SF rates are triggered, that would often consume all the gas content within less than a billion years, is not yet known. In most SBDG large HI reservoirs enveloping the luminous galactic body have been recently detected (NGC 4449: Hunter et al. 1998, I Zw 18: van Zee et al. 1998b, NGC 1705: Meurer et al. 1998) and can to some extent be interpreted to fall in and, by this, to feed the SB (e.g. He 2-10: Kobulnicky et al. 1995). On the other hand, because of their low binding energy SBDGs are characterized by superwinds (Marlowe et al. 1995) or large expanding X-ray plumes which are driven by SNeII. Their existence demonstrates the occurrence of metal loss by means of large-scale galactic winds and allows to study this cosmologically common phenomenon still today.

Even more smoothly developing dIrrs have experienced several SF epochs, like the LMC (Pagel & Tautvaisiene 1998). Their irregular shape, however, stems from their present mainly patchy distribution of HII regions like it is visible e.g. in some prominent representatives NGC 4449 and NGC 4214. Although dIrrs consist of the same and higher gas fraction as giant spiral galaxies (gSs), they appear with a wide range but lower metallicity $Z$ as gSs. Again this implies that the metal-enriched gas from SNeII was lost from the galaxy. Vice versa also infall of low-metallicity (even primordial) gas can allow for this signature, but reasonably both might have acted. Nevertheless, very metal-poor dIrrs like e.g. I Zw 18 (Searle & Sargent 1972), GR8 (Skillman et al. 1988), or SBS 0335-52 (Thuan et al. 1997) seem to form their first generation of stars today. In addition, a strange but perhaps common mode of DG formation exists in tidal tails of merging galaxies (Mirabel et al. 1992, Duc & Mirabel 1998). Although these units seem to be gravitationally bound, their survival of energetic events like accumulative SNeII or tidal friction is questionable.

## 1.2. Element Abundances and the Chemical Evolution of Dwarf Irregulars

While the low metal content in DGs can be attributed to both metal-enhanced mass loss and/or low-metallicity gas infall, another puzzling fact needs explanation: Why do DGs, though with O abundances below 1/10 solar, show also low N/O ratios of up to 0.7 dex smaller than in gSs with a large scatter and no significant correlation with O/H (Pagel 1985; see fig.1)? It is persuasive that normal disk HII regions and stars are located along the line of secondary nitrogen production (SNP), while the N/O ratios of DGs scatter around an average value of -1.5 independently of their oxygen abundance. This is supposed to represent the primary production process of N (PNP) at low metallicities. Most interestingly, van Zee et al. (1998a) have recently found that the abundances of the outermost HII regions in spiral galaxies at large radii have not only low O abundances, but spread over the same regime between SNP and PNP tracks and overlap with DG's values in the sense that they approach the primary line (fig.2).

For a further insight and deeper understanding of stellar yields and galactic element enrichment at early galactic stages one can compare N/O vs. O measurements of extragalactic objects from different cosmological ages. Abundances for quasar emission lines reach to very high abundances of one order of magnitude above solar (Hamann & Ferland 1992) as an extrapolation of the SNP line, while QSO absorption-line systems (ALS) scatter again without any strong tendency except that damped-Ly$\alpha$ systems (DLAs) (Molaro et al. 1996, 1998, Pettini et al. 1995) are more widely distributed than Lyman-limit systems (LLS) (Köhler et al. 1996) (see fig.3).

Past standard stellar evolution models yield the following separation: While C and N are attributed by planetary nebulae (PNe) from intermediate-mass stars (IMS) to the warm cloudy gas phase (CM) of the interstellar medium (ISM), O and Fe are dominantly released by means of SNII and SNIa explosions, respectively, to the hottest phase of the ISM, the intercloud medium (ICM) and, therefore, trace these energetic events and their stellar progenitors. This differentiation of N and O pollution to the gas phases has led several authors to propose various chemical models which could successfully reach the observed N/O-O regime of DGs. Matteucci & Tosi (1985), Marconi et al. (1994) and Kunth et al. (1995) allow for galactic winds with selective element depletion, which are driven by numerous SB episodes, in order to account for the necessary reduction of O. Similarly, Garnett (1990) and Pilyugin (1992) present models that follow the assumption of abundance self-enrichment within the observed HII regions as it was concluded before from observations (Kunth & Sargent 1986).

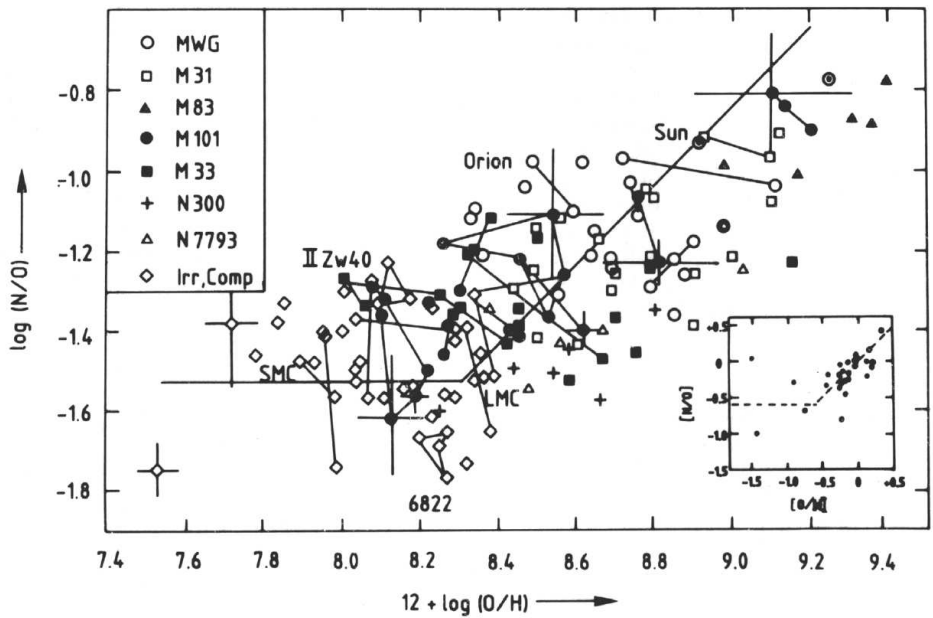

Figure 1.   N/O ratio over the oxygen abundance of HII regions in dwarf irregular and spiral galaxies including the Milky Way and the Sun. The lines represent the averaged tendency of halo and disk stars in the MWG. [taken from Matteucci (1986) who has completed the original figure by Pagel (1985) with the inlay from Tomkin & Lambert (1984) on disk and halo stars.]

All these models differentiate the O and N release due to discontinuous and separate events in accordance to their progenitor lifetimes, by this, leading to saw tooth-like evolutionary tracks in the log(N/O)-log(O/H) diagram (Pilyugin 1992). Parameters like wind efficiency, IMF, and initial gas fraction provide sufficiently good model approaches to the observations. It should be emphasized that three important ingredients enter mainly these referred chemical evolutionary models: 1) The yields are in principle based on almost the same stellar evolution calculations (N by Renzini & Voli (1981) and O by Woosley & Weaver (1986) or by Maeder (1992) and Maeder & Meynet (1989)). 2) All the investi-gations start already from N/O values (even at low O abundance) reached by the element enrichment of a normal stellar population with traditional yields. 3) Gas infall of metal-poor gas is not considered because it reduces only the oxygen abundance while leaving N/O almost constant.

Figure 2.     N/O ratio over the oxygen abundance of Hɪɪ regions in dwarf irregular and spiral galaxies (from van Zee et al. 1998a). The transition from low-metallicity Hɪɪ regions at large radii of spirals to dIrrs is discernible.

If N and O are produced by different progenitor stars and are separately released to the different phases of the ISM, their simultaneous existence in Hɪɪ regions representing the ionized CM, could only result from mixing processes between CM and ICM. In models where these processes are taken into account, the observed N/O ratio should, therefore, permit qualitative studies on both the mixing direction and efficiency. Additionally, its radial distribution in a DG provides an insight into dynamical effects of the ISM. The assumption of an individual self-enrichment of Hɪɪ regions during their ionization-caused observability (Pilyugin 1992) would plausibly involve both, abundance variations during their visibility epoch and, by this, differences between Hɪɪ regions within the same dIrr. In contrast, Kobulnicky et al. (1999) report the non-detection of any sizable O, N, and He anomalies from Hɪɪ regions in the vicinity of young star clusters in SBDGs with one exception, NGC 5253, what reveals a central N overabundance (Kobulnicky et al. 1997). In NGC 1569, that has recently formed two SSCs, Kobulnicky & Skillman (1997) find constant O and N/O values over a radial extent of more than 600 pc with a scatter of N/O by only 0.2 dex, while self-enrichment tracks reach nearly one dex in N/O (Pilyugin 1992) and dispersal distances of much less. The same appearance is found in I Zw 18 on a scale of 1 kpc (Izotov 1999). From a study of five dIrrs Kobulnicky & Skillman

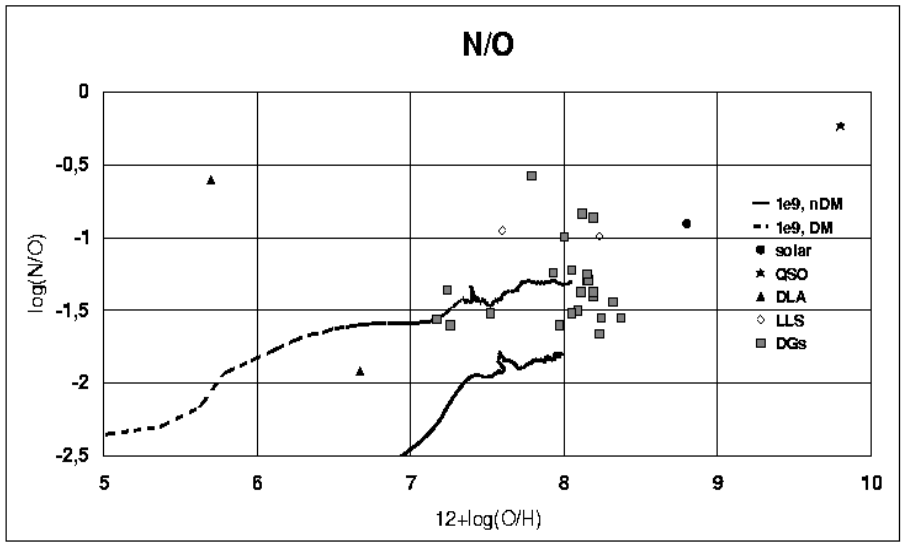

Figure 3.    Evolutionary tracks of 2d chemodynamical models for $10^9 \, \mathrm{M}_\odot$ galaxies with (upper dashed curve) and without DM halo (lower full line) in comparison with N/O vs. O/H measurements of H<span>II</span> regions in a few dIrrs (squares), of one typical QSO (asterix), two damped-Ly$\alpha$ systems (triangles) and two Lyman-limit systems (diamonds). For more information see text.

(1998) stated that the hypothesis of differential mass (and by this element) loss is inconclusive.

## 2.    Chemodynamical Models

Although many authors concentrate on the more spectacular SBDGs, not all observed dIrrs that fill in the same N/O-O regime have passed or experience at present strong SBs, but form stars also at an almost moderate and continuous rate. Empirical studies and theoretical investigations under galaxy conditions of low gravitational potentials have proven that their ISM should be balanced by counteracting processes like SN heating and cooling, turbulence and dissipation (Burkert & Hensler 1989). Similarly, the SF itself is self-regulated under various conditions due to stellar heating, gas cooling, evaporation, and condensation (see e.g. Franco & Cox 1983, Köppen et al. 1995, 1998).

If the evolution of galaxies is sensitive to the energetical impact from different processes, an adequate treatment of the dynamics of stellar and gaseous components and of their mutual energetical and materialistic interactions is essential. This modeling of galactic evolution is properly treated by the chemo-dynamical (*cd*) prescription. Its formulation in 1d and 2d dynamics with the "materialistic" and "energetic" equations can be found for interested readers e.g. in Theis, Burkert, & Hensler (1992) and Samland, Hensler, & Theis (1997), respectively. The evolution of DGs can proceed in self-regulated ways both, globally by large-scale flows of unbound gas but also locally in the SF sites. *cd* models of non-rotating DG systems achieve SBs from the initial collapse and after mass-loss induced expansion in the recollapse phase of its bound gas fraction (Hensler et al. 1993, 1999). Particular models are evolving also with oscillatory SF episodes. The formed stellar populations are similar to dSphs as well as more massive BCDGs and reveal a central concentration of the recent SF region embedded in older elliptical stellar populations. External effects (Vilchez 1995) like extended dark matter (DM) halos, the intergalactic gas pressure, etc., could cause further morphological differences of DGs e.g. by regulating the outflow and allowing for a recollapse or even fuelling subsequent SF by infall of intergalactic clouds.

For rotating dIrrs *cd* models are essential which account at least for a 2d rotationally symmetric structure. 3d calculations (in progress) have to make use of an adaptive-mesh refinement algorithm for numerical reasons in order to find a compromise between acceptable computer time and sufficient spatial resolution. In our models a gaseous protogalactic cloud starts with or without a DM halo (Burkert 1995) and with a Plummer-Kuzmin-type gas distribution. In the following we wish to discuss rotating $10^9\,M_\odot$ DG models which commence with a 2 kpc baryonic density scalelength. Recent stellar evolution models by Woosley & Weaver (1995), which provide a SNP (and a small amount of PNP) also by HMS are taken into account. For consistency also the most recent models for IMS by van der Hoek & Groenewegen (1997) have been applied.

The evolutionary model of two $10^9\,M_\odot$ models (with and without a DM halo) are shown in fig.3. Here the abundances of N and O are determined as the mean value in bright SF regions of the models in order to achieve the best comparability with observations of HII regions. This new deduction of abundances differs from the already published model results (Hensler & Rieschick 1998, 1999) in the sense that the former abundances have been calculated as the mass-weighted mean of the whole computational domain, i.e. even with the halo gas, and all components. Although these former results provided already a first hint that the abundance ratios seem to deviate from simple models, they fail to be representative for the visible part. The evolutionary states of the models in fig.3 at the upper points of the tracks are shown at ages of 3.8 Gyrs for the DM model and of 3.4 Gyrs for the other without DM, respectively.

Our *cd* models reveal three major and significant discrepancies to former above-mentioned chemical models:
1) Like normal dIrrs and analytical studies show, our models do not evolve into a global SB, but only to local SF enhancements (Rieschick & Hensler 1999). This leads to the conclusion, that SBs might need a trigger mechanism as observed in most SBDGs (see section 1.1).
2) In contrast to the metal enrichment in closed-box models, here the evolutionary track in the N/O-O diagram does not circumvent the regime of the observed values for dIrrs along a horizontal track at almost solar N/O ratios, but passes from low N/O and O/H values in the diagonal direction between PNP and SNP trends through the regime of dIrr observations (fig.3).

Figure 4.    Evaporation/condensation parameter $\beta$ in grey colors at 0.1 (upper left), 0.2, 0.3 (lower left), and 1.5 Gyrs after the onset of the model simulation. The darkest region at the lower left corner (center of the galaxy) of each panel represents $\beta = -1.$, i.e. complete evaporation, while in the upper right brighter part $\beta$ amounts to almost $+0.2$. For the definition of $\beta$ see text.

3) From fig.4 it becomes visible that N mixes perfectly with O in the ICM due to evaporation of the CM in the vicinity of the SF and SNII explosion sites. This achieves a nearly constant abundance ratio widely spread over large distances within the DG. The evaporation/condensation parameter, that is defined as $\beta = (\rho_{cond} - \rho_{evap})/(\rho_{cond} + \rho_{evap})$, reachs -1.0 in the innermost center, while condensation starts to dominate outward of the steeper density gradient (also recognizable as $\beta$ gradient) where $\beta$ can amount to +0.2.

While this latter fact lends supports to the observed radial constancy of the N and O abundances in a BCDGs, by this, it conflicts with the self-pollution argument for the H$_{II}$ region except in the very center. In addtion, it contradicts to the selective expulsion of O by means of galactic winds (Matteucci & Tosi 1985). Obviously, a substantial mixing of CM and ICM determines the chemical evolution. To summarize, the dIrr abundance problem cannot be solved by SB-driven galactic wind, because the N/O ratio should remain constant.

### 3. Gas Infall

Generally speaking and as a first still careful interpretation of the presented chemical evolution of *cd* dIrr models, one has to study the reasons for the occuring discrepancies of N/O abundance ratios between simple models and the *cd* models which, on the other hand, agree with DG observations. Although even with the new additional nitrogen production from HMS the shortly living IMS of around 5-8 M$_\odot$ should release their primary N on sufficiently short timescales and rise N/O to solar values. As a first step one has to compare the dynamical "*cd*" models with closed-box results under the inclusion of the same interaction prescriptions. Although the evaporation/condensation behaviour of the closed-box evolution might differ from the dynamical model and, by this, the element abundances in the gas phases should deviate, the global box model, i.e. the sum of all gaseous components, should gain the same abundance $Z$ in accordance to the analytical solution. Without going into too much detail here we wish to emphasize this agreement with the expected analytical behaviour and refer to a forthcoming paper (Rieschick & Hensler 1999). We conclude that this issue excludes the possible explanation that the discrepancy stems solely from the application of new stellar yields. More preferably, it must be caused by the dynamical behaviour of gas in a low gravitational potential with a high energy impact. Only if the SF rate is increasing steeply, the SNP of HMS can exceed the delayed release of primary N by IMS.

Recently, Köppen & Edmunds (1999) have performed analytical studies of the effect of gas infall into a star-forming galaxy. Since the infalling gas is assumed to be primordial, it reduces the normal stellar yields in the sense that it decelerates the temporal metal enrichment or it approaches an equilibrium value. Formally, a constant ratio a=A/Ψ of temporal gas infall rate A(t) normalized to the star-formation rate Ψ(t) yields a steadily increasing evolutionary path in the diagram joining purely secondary-to-primary element abundances $Z_s/Z_p$ vs. primary element. Also monotonically increasing or decreasing a(t) evolve along

Figure 5. Schematic global evolution of the abundance ratio of secondary-to-primary elements $Z_s/Z_p$ vs. primary element content normalized to its stellar yield $y$.

The left panel shows the arbitrarily chosen temporal gas-infall rate A normalized to the star-formation rate $\Psi$ as a=A/$\Psi$ over the increasing stellar mass fraction (after Köppen & Edmunds 1999). In the right panel the general element abundances are shown as the ratio of purely secondary-to-primary elements, $Z_s$ and $Z_p$, respectively, what can be assumed to be representative for nitrogen and oxygen production by massive stars. The thick diagonal line correspond to the $Z_s/Z_p - Z_p$ evolution of an a=0. model.

The figures have been generated by application of the applet by Köppen (1998). For discussion of the curves see text.

the secondary-production track, while for da/dt>0 an equilibrium point in this diagram will be reached at lower value than for constant or decreasing a(t). In a situation where significant gas infall occurs in a stochastic event, the enrichment path in the diagram makes a loop to lower $Z_p$ but turns back to the former track at lower $Z_s/Z_p$ when a(t) drops again to a state as before the infall episode. Even the beginning from an arbitrary point in this diagram cannot prevent to converge to the secondary production track again. While the formalism is described in their paper (Köppen & Edmunds 1999), the attention of interested readers should be drawn to an appropriate and instructive applet by one of us (Köppen 1998), where the analytical solutions of the set of equations can be comfortably visualized for different situations and different possible combination of variables.

In fig.5 we wish to demonstrate how the observed scattered values of N/O with oxygen in dIrrs could be interpreted: After an initial infall epoch which represents the early collapse phase of the gaseous DG, a later episodic infall event can deflect the evolutionary track for the ratio of a purely secondarily produced element to a primary one towards lower single abundance at the same ratio. As can be discerned, the loop can pass through the regime of observed dIrr measurements. Our models can start from a basis of SNP with proceedingly enhanced PNP but additional gas infall episodes. The slope of the evolutionary path is then supposed to wobble between secondary and horizontal slope and because of small but varying gas infall rates. The chemical evolution of the *cd* models is directly and self-consistently passing through the regime of abundances observed in dIrrs (see fig.3).

## 4. Conclusions

In this paper we have presented recent *cd* models of dIrrs that are aimed to investigate the reason of their observed peculiar element abundances and, by this, to address the question on their chemical evolution. Because of their capability to treat gas and stars dynamically within the framework of their interactions the self-consistent *cd* models discovered strong constraints on the assumptions of analytical considerations. In particular, mixing effects of the gas phases prevent a selective element depletion, and furthermore, the models evolve with local variations but globally moderate SF activity (Rieschick & Hensler 1999) and therefore typically for the major fraction of dIrrs. For actuality, the most appropriate yields from recent stellar evolution models for HMS (Woosley & Weaver 1995) and IMS (van der Hoek & Groenewegen 1997) are implied, where nitrogen is secondarily and (to less extent) primarily produced by HMS. Nevertheless, since the PNP in IMS raises reasonably the N/O ratio within a typical timescale of their stellar lifetimes to almost solar, i.e. larger than observed in dIrrs, a rapid O enhancement could be required. This invokes a present SB event as a plausible explanation.

As a perspective, the observed low N/O ratios in DGs' HII regions, however, could be attained, if N would be purely produced as a secondary element in HMS. Nevertheless it is unavoidable, that the PNP by IMS should also dominate in a single stellar population after around $10^8$ yrs and therefore enhances the N/O ratio. In contrast to the mentioned expectations, the chemical evolutionary track of the here presented *cd* model is surprising as it passes through the regime of observed abundances in dIrrs without any artificial manipulation. Three mechanisms could be considered for an explanation of this issue. Firstly, N and O are almost perfectly mixed in the innermost regions of dIrrs due to overwhelming evaporation of clouds embedded in the hot ISM and, by this, not allowing for a selective element loss in dIrrs. Although only a weak galactic wind might be produced in self-regulated low-mass galaxies, i.e. without any SB the centrally peaked SF drives a large-scale radial redistribution of N/O according to the presently exploding HMS in a surrounding ISM which is not yet completely polluted by primary N from IMS because of their longer lifetimes. The lower N/O hot gas mixture condensates at remote regions where it is also not in accordance with the locally existing stellar population. This issue involves that dIrrs are systems still forming from their large surrounding gas reservoir or strongly affected by it. Secondly, the *cd* results can be interpreted with respect to further gas infall in a global model where the SNP in HMS exceeds the PNP in IMS. For the first billion years this could be the case for a steeply increasing infall rate where the rapidly released HMS yields exceed the delayed PNP. At least, analytical studies show that just this latter situation can be reached repeatedly if gas infall episodes occur. By such events of different strength and durations the whole regime of secondary-to-primary element ratios in dIrrs can be covered.

Since recent deep observations of dIrrs and SBDGs have discovered an increasing existence of large gas envelops, i.e. of enormous gas reservoirs circling or accumulating around the visible body of the BCDGs, the relevance of infall episodes for the evolution of these low-mass galaxies is doubtless and serves as the most promising explanation for their observed abundances and their rejuvenation. Therefore, this process is essential for evolutionary models. Since the SF can be triggered subsequently by such gas infall, SBs are a plausible consequence. Although the smooth collapse of a protogalactic cloud like in our conditions resembles somehow a steady, but decreasing gas infall rate, stochastic infall of intergalactic gas clouds is not yet included in the presented models. As a preliminary result from this discussion one should, however, conclude that their effect on the abundances is not mainly attributed to a galactic wind but to the gas infall. This could considerably amplify the tendency found in the presented *cd* results and lead to repeated element enhancement cycles in the N/O-O diagram.

As a great success the *cd* models reproduce already convincingly the observed abundance peculiarities of dIrr in a self-consistent way. Since the *cd* treatment is appropriate for sensitively balanced systems of low gravitational potential energy like the disks of gSs and dIrrs, its models can provide a fundamental insight into interaction processes of small-scale energetics and large-scale dynamics. This has already led to a successful reproduction of metallicity distributions and the temporally coupled evolution of different regions in the Milky Way (Samland et al. 1997) by means of large-range streaming of hot gas and the mixing with locally infalling gas clouds. This same interaction of processes seems to affect also the evolution of dIrrs substantially.

Acknowledgments: We are grateful to Mike Edmunds for helpful discussions. This work was supported by the *Deutsche Forschungsgemeinschaft* (DFG) under grant no. He 1487/5-3 and /23-1 (A.R.). The numerical calculation have been performed at the computer center of the University of Kiel and the NIC in Jülich.